\documentstyle[conference,a4wide,amsfonts,cite,epsfig]{IEEEtran}

\newcommand{\vb}[1]{{\mathbf{#1}}}
\newcommand{\lb}[1]{\label{#1}}

\newcommand{\bc}{\begin{center}}
\newcommand{\ec}{\end{center}}
\newcommand{\be}{\begin{equation}}
\newcommand{\ee}{\end{equation}}
\newcommand{\bea}{\begin{eqnarray}}
\newcommand{\eea}{\end{eqnarray}}
\newcommand{\ba}[1]{\begin{array}{#1}}
\newcommand{\ea}{\end{array}}

\newcommand{\bt}[1]{\begin{table}[ht]\centering\begin{tabular}{#1}}
\newcommand{\et}[1]{\end{tabular}\caption{\small#1}\end{table}}

\begin{document}

\title{Pseudo-Photon Radiative Corrections in Non-Regular Backgrounds}
\author{\authorblockN{P. Castelo Ferreira}\\[2mm]
\authorblockA{CENTRA, Instituto Superior T\'ecnico,\\ Av. Rovisco Pais,\\ 1049-001 Lisboa, Portugal\\[2mm]
e-mail: \tt pedro.castelo.ferreira@ist.utl.pt}}

\maketitle

\begin{abstract}
Extended $U_e(1)\times U_g(1)$ electromagnetism
containing both a photon and a pseudo-photon
is introduced at variational level and
is justified by the violation of the Bianchi
identities in conceptual systems, either in the presence of magnetic
monopoles or non-regular external fields,
not being accounted by the standard Maxwell
action. It is shown that in the perturbative
quantum field theory regimes there are observable
consequences both for the magnetic momenta and
second order vacuum polarization radiative corrections
in the presence of non-regular strong electromagnetic
fields (e.g. rotating magnetic fields).\\[2mm]
\noindent PACS: 03.50.De, 12.20.-m,11.15.-q\\
Keywords:  pseudo-photon, vacuum polarization, magnetic moment, Euler-Heisenberg\\
\end{abstract}

\section{Maxwell Action}

The Maxwell equations where derived phenomenologically and unified in 1861~\cite{Maxwell}
\be
\ba{rcl}\displaystyle\bf{\nabla}\cdot\bf{E}&=&\displaystyle\rho_e\\[3mm]
			                         \displaystyle\bf{\nabla}\times\bf{B}-\dot{\bf{E}}&=&\displaystyle{\bf{j}}_e\\[3mm]
			                         \displaystyle\bf{\nabla}\cdot\bf{B}&=&\displaystyle 0\\[3mm]
			                         \displaystyle\bf{\nabla}\times\bf{E}+\dot{\bf{B}}&=&\displaystyle 0\ea
\lb{eq_Maxwell}
\ee
At variational level these equations are described by the well known Maxwell action
\be
S_{\mathrm{Maxwell}}=-\int d^4x \left[\frac{1}{4}F_{\mu\nu}F^{\mu\nu}+A_\nu J^\nu_e\right]
\lb{S_Maxwell}
\ee
with the usual definition of the gauge connection $F_{\mu\nu}=\partial_\mu A_\nu-\partial_\nu A_\mu$,
being the 2 massless degrees of freedom of the photon encoded in the vector gauge field $A_\mu$.
In the following we are using the standard definitions of the electromagnetic fields
\be
E^i=F^{0i}\ \ ,\ \ \ B^{i}=\frac{1}{2}\epsilon^{ijk}F_{jk}
\ee
The equations of motion are obtained by considering a functional derivation of the action with
respect to the gauge field $A_\mu$
\be
\ba{rcl}
\displaystyle\frac{\delta S}{\delta A_\mu}=0&\Leftrightarrow&\displaystyle\partial_\mu F^{\mu\nu}=J_e^\nu\\[2mm]
 & \Rightarrow&\displaystyle\left\{\ba{rcl}\displaystyle\bf{\nabla}\cdot\bf{E}&=&\displaystyle\rho_e\\[3mm] \displaystyle\bf{\nabla}\times\bf{B}-\dot{\bf{E}}&=&\displaystyle{\bf{j}}_e\ea\right.
\ea
\ee
and as can explicitly checked by direct comparison with the original Maxwell equations~(\ref{eq_Maxwell})
only hold half of the these equations. The remaining equations are obtained by demanding (or imposing)
regularity of the gauge fields. This requirement is translated into the Bianchi identities
\be
\epsilon^{\mu\nu\lambda\rho}\partial_\nu F_{\lambda\rho}=0\,\Rightarrow\, \left\{\ba{r}\displaystyle\bf{\nabla}\cdot\bf{B}= 0\\[3mm]
			                         \displaystyle\bf{\nabla}\times\bf{E}+\dot{\bf{B}}= 0\ea\right.
\ee
which reproduce the remaining Maxwell equations as can be checked by direct comparison with~(\ref{eq_Maxwell}).

Let us enumerate some important remarks concerning the Maxwell action~(\ref{S_Maxwell}):
\begin{enumerate}
\item is a successful functional principle in most fields of physics dealing with electromagnetic interaction, both at classical and quantum level;
\item is the first example of fundamental interactions unification based in the phenomenologically derived Maxwell equations;
\item is in the basis for todays particle physics, in particular the standard model;
\item however does not reproduce the Bianchi Identities which are imposed externally, hence for systems where violation of these identities is relevant fails to describe the full Maxwell equations at variational level.
\end{enumerate}
The last point clearly holds a problem when trying to describe some particular conceptual physical systems.
Examples of such systems are the description at variational level of magnetic monopoles and external non-regular fields.
The existence of Magnetic monopoles are the only theoretical justification for the quantization of electric charge~\cite{Dirac,Jackson},
however have so far not been experimentally detected and most probably due to the Dirac quantization condition expressed in
terms of the unit electric charge $e$ and magnetic charge $g$
\be
eg=2\pi\hbar\,n\ \ ,\ \ n\in\mathbb{N}\ ,
\lb{Dirac_cond}
\ee
are in a strong coupling regime, hence confined~\cite{Nambu}. Therefore free magnetic charges can only be considered at theoretical level.
Also non-regular electromagnetic fields are usually present only in conceptual systems (except probably for type~II superconductivity)
where extended singular objects or topological defects are present, for example the Abrikosov-Nielsen-Olesen string~\cite{strings}.

\subsection{Magnetic Monopoles}

The inclusion of magnetic currents $J_g^\mu=(\rho_g,j_g^i)$ directly in the Maxwell equations is straight forward~\cite{Jackson}
\be \left.\ba{r}\displaystyle\bf{\nabla}\cdot\bf{B}= \rho_g\\[3mm]
			                         \displaystyle\bf{\nabla}\times\bf{E}+\dot{\bf{B}}= \vb{j}_g\ea\right\}
\,\Rightarrow\,\epsilon^{\mu\nu\lambda\rho}\partial_\nu F_{\lambda\rho}=J_g^\mu
\ee
however these expressions clearly imply a violation of the Bianchi identities. Furthermore:
\begin{itemize}
\item these equations are not deducible from the Maxwell action;
\item implies extended singularities, either the Dirac string~\cite{Dirac} or the Wu-Yang fiber bundle~\cite{WY}.
\end{itemize}
So we conclude both that in the presence of magnetic monopoles the Maxwell action~(\ref{S_Maxwell}) is an
incomplete description of electromagnetic interactions and that it implies the existence of infinite singularities
which can hardly be considered real physical objects.

\subsection{Non-Regular External Electromagnetic Fields}

Non-regular electromagnetic fields consist of field configurations with either a time-dependent
magnetic field or a non-vanishing rotational of the electric field
\be
\left.\ba{rcl}\displaystyle\dot{\bf{B}}&\neq&0\\[3mm]\displaystyle\bf{\nabla}\times\bf{E}&\neq&0\ea\right\}\ \Rightarrow\ \bf{\nabla}\times\bf{E}+\dot{\bf{B}}\neq 0
\ee
which again:
\begin{itemize}
\item violates the Bianchi identities;
\item is not deducible from the Maxwell action.
\end{itemize}

These two examples clearly show that a full description of electromagnetic interactions is not
achieved by the Maxwell action. Next we show that a possible solution for this problem is to
consider an extended Abelian gauge group $U_e(1)\times U_g(1)$, i.e. in
addition to the standard \textit{electric} vector gauge field (photon) to consider a \textit{magnetic}
pseudo-vector gauge field (pseudo-photon).

\section{Extended $U_e(1)\times U_g(1)$ Electromagnetism}

Let us consider two gauge fields, a vector field $A$ (the standard photon) and a
pseudo-vector field $C$ (the pseudo-photon). We will take the following assumptions:
\begin{enumerate}
\item $P$ and $T$ invariance of electromagnetic interactions;
\item existence of only one electric and one magnetic physical fields.
\end{enumerate}
Both this assumptions are justified experimentally. From the Maxwell equations
the magnetic currents transform as pseudo currents, hence considering minimal
coupling of these currents to the $C$ field implies it also to transform as
a pseudo-vector in order to ensure $P$ and $T$ invariance at variational level.
Given the above assumptions, up to a sign choice of the topological term, we get
the only possible action~\cite{CF,Sing,EB}
\be
\ba{rcl}S&=&\displaystyle \frac{1}{4}\int d^4x\left[-F_{\mu\nu}F^{\mu\nu}+G_{\mu\nu}G^{\mu\nu}\right.\\[3mm]
         & &\displaystyle \ \ \ \ \ \ \ \ \ \ \ \ +\epsilon^{\mu\nu\lambda\rho}F_{\mu\nu}G_{\lambda\rho}\\[3mm]
                        & &\displaystyle \ \ \ \ \ \ \ \ \ \ \ \ +e\,\left(A_\mu-\tilde{C}_\mu\right)J_e^\mu\\[3mm]
                        & &\displaystyle \ \ \ \ \ \ \ \ \ \ \ \ \left.-g\,\left(\tilde{A}_\mu-C_\mu\right)J_g^\mu\ \ \right]\ea
\lb{S_AC}
\ee
in terms of the gauge connections $F_{\mu\nu}=\partial_\mu A_\nu-\partial_\nu A_\mu$ and
$G_{\mu\nu}=\partial_\mu C_\nu-\partial_\nu C_\mu$ and with the electromagnetic field definitions
\be
\ba{rcl}
E^i&=&\displaystyle F^{0i}-\frac{1}{2}\epsilon^{0ijk}G_{jk}\\[2mm]
B^i&=&\displaystyle G^{0i}+\frac{1}{2}\epsilon^{0ijk}F_{jk}
\ea
\lb{EB_fields}
\ee
The couplings to the electric and magnetic currents are derived by considering the Lorentz force
expressions in terms of these definitions for the electromagnetic fields (see~\cite{EB} for more details).
The tilde fields do not constitute new degrees of freedom and are defined in terms of the differential
equations
\be
\tilde{F}^{\mu\nu}=\frac{1}{2}\epsilon^{\mu\nu\lambda\rho}F_{\lambda\rho}\ ,\ \ \tilde{G}^{\mu\nu}=\frac{1}{2}\epsilon^{\mu\nu\lambda\rho}G_{\lambda\rho}
\ee
with $\tilde{F}_{\mu\nu}=\partial_\mu \tilde{A}_\nu-\partial_\nu \tilde{A}_\mu$ and $\tilde{G}_{\mu\nu}=\partial_\mu \tilde{C}_\nu-\partial_\nu \tilde{C}_\mu$. In addition we note that the coupling constants $e$ and $g$ correspond
to the electric and magnetic unit charges and obey Dirac's condition~(\ref{Dirac_cond})

The equations of motion for the $A$ and $C$ field now reproduce the full Maxwell equations in
terms of the above electromagnetic field definitions~(\ref{EB_fields})
\be
\ba{rcl}
\displaystyle\frac{\delta S}{\delta A_\mu}=0&\Leftrightarrow&\displaystyle\partial_\mu \left(F^{\mu\nu}-\frac{1}{2}\epsilon^{\mu\nu\lambda\rho}G_{\lambda\rho}\right)=J_e^\nu\\[3mm]
&\Rightarrow&\displaystyle\left\{\ba{rcl}\displaystyle\bf{\nabla}\cdot\bf{E}&=&\displaystyle\rho_e\\[3mm] \displaystyle\bf{\nabla}\times\bf{B}-\dot{\bf{E}}&=&\displaystyle{\bf{j}}_e\ea\right.\\[8mm]
\displaystyle\frac{\delta S}{\delta C_\mu}=0&\Leftrightarrow&\displaystyle\partial_\mu \left(G^{\mu\nu}+\frac{1}{2}\epsilon^{\mu\nu\lambda\rho}F_{\lambda\rho}\right)=J_g^\nu\\[3mm]
 &\Rightarrow&\displaystyle\left\{\ba{rcl}\displaystyle\bf{\nabla}\cdot\bf{B}&=&\displaystyle \rho_g\\[3mm]
			                         \displaystyle\bf{\nabla}\times\bf{E}+\dot{\bf{B}}&=&\displaystyle {\bf{j}}_g\ea\right.
\ea
\ee
We note that for regular gauge field configurations the Bianchi identities for $A$ and $C$ hold
and the equations decouple. This is no longer the case for non-regular external fields.
In order to see it explicitly let us consider a decomposition of the vector gauge field $A_\mu=\bar{A}_\mu+a_\mu$
into internal $a$ and external $\bar{A}$ components such that in terms of the gauge connections $F_{\mu\nu}=\bar{F}_{\mu\nu}+f_{\mu\nu}$ the equations of motion read
\be
\ba{c}
\displaystyle\partial_\mu \left(f^{\mu\nu}-\frac{1}{2}\epsilon^{\mu\nu\lambda\rho}G_{\lambda\rho}\right)+\partial_\mu\bar{F}_{\mu\nu}=0\\ \Downarrow\\[3mm]
\displaystyle \left\{\ba{r}\displaystyle\bf{\nabla}\cdot\bf{E}+\bf{\nabla}\cdot\bf{E}^{\mathrm{ext}}= 0\\[3mm]
\displaystyle\bf{\nabla}\times\bf{B}-\dot{\bf{E}}+\bf{\nabla}\times\bf{B}^{\mathrm{ext}}-\dot{\bf{E}}^{\mathrm{ext}}= 0\ea\right.\\[10mm]
\displaystyle\left(G^{\mu\nu}+\frac{1}{2}\epsilon^{\mu\nu\lambda\rho}f_{\lambda\rho}\right)+\frac{1}{2}\epsilon^{\mu\nu\lambda\rho}\partial_\nu \bar{F}_{\lambda\rho}=0\\
\Downarrow\\[3mm]
\displaystyle \left\{\ba{r}\displaystyle\bf{\nabla}\cdot\bf{B}+\bf{\nabla}\cdot\bf{B}^{\mathrm{ext}}= 0\\[3mm] \displaystyle\bf{\nabla}\times\bf{E}+\dot{\bf{B}}+\bf{\nabla}\times\bf{E}^{\mathrm{ext}}+\dot{\bf{B}}^{\mathrm{ext}}= 0\ea\right.
\ea
\lb{Maxwell_ext}
\ee

So extended $U_e(1)\times U_g(1)$ electromagnetism:
\begin{enumerate}
\item has at classical level the same results expressed by the Maxwell equations~(\ref{eq_Maxwell}). This is
due to the definition of the electromagnetic fields in~(\ref{EB_fields});
\item solves the inconsistency concerning the violation of the Bianchi identities at variational level in the
presence of either monopoles or non-regular field configurations.
\end{enumerate}
As we will show next at quantum field theory level there will be observable consequences of the second gauge sector,
namely for:
\begin{itemize}
\item magnetic momenta of fermions;
\item second order vacuum polarization.
\end{itemize}
Also it is important to stress that at quantum level the pseudo photon is a ghost (or phantom) due to the
opposite sign of the kinetic term in~(\ref{S_AC}), however in the framework we are considering
this does not pose any problem since the $C$ field is only present as a background field and is not
quantized.

\section{Radiative Corrections}

In order to couple the gauge fields to fermions we consider the usual Dirac Lagrangian
\be
{\mathcal{L}}_\psi=\bar{\psi}\left(i\gamma^\mu\partial_\mu-e\gamma^\mu A_\mu-g\gamma^\mu\tilde{C}_\mu-m\right)\psi
\ee
Recalling that $e$ and $g$ are related by Dirac's quantization condition~(\ref{Dirac_cond}) we obtain, for $n=1$ the
dimensionless relation between the unit charges and the fine-structure coupling constants
\be
\ba{rcl}
\displaystyle\sqrt{\frac{\epsilon_0}{\mu_0}}\,\frac{g}{e}&\approx&68.5\\[3mm]
\displaystyle\frac{\alpha_g}{\alpha_e}&=&\displaystyle\frac{e^2}{4\pi\epsilon_0\hbar c}\,/\,\frac{g^2}{4\pi\mu_0\hbar c}\approx 4692.2
\ea
\ee
These results will have observable results both for the magnetic momenta and vacuum polarization
effects in non-regular electromagnetic background fields. Let us consider the most simple example
for such conceptual backgrounds, a rotating magnetic field in vacuum~\cite{pseudo}
\be
\ba{l}
\displaystyle\vb{B}^{\mathrm{ext}}(t)= B_0\left[\sin(\omega_0\,t),\cos(\omega_0\,t),0\right]\\[3mm]
\displaystyle\dot{\vb{B}}^{\mathrm{ext}}(t)=\omega_0\,B_0\left[\cos(\omega_0\,t),-\sin(\omega_0\,t),0\right]
\ea
\ee
The Maxwell equations~(\ref{Maxwell_ext}) can be solved recursively (see~\cite{pseudo} for further details)
such that we obtained the induced fields
\be
\ba{l}
\displaystyle\vb{E}^{\mathrm{ind}}= B_0\left[0,0,-\sin(\omega_0\,y)\cos(\omega_0\,t)\right.\\[2mm]
\displaystyle\left.\ \ \ \ \ \ \ \ \ \ \ \ -\sin(\omega_0\,x)\sin(\omega_0\,t)\right]\\[3mm]
\displaystyle\vb{B}^{\mathrm{ind}}= B_0\left[(\cos(\omega_0\,y)-1)\sin(\omega_0\,t),\right.\\[2mm]
\displaystyle\left.\ \ \ \ \ \ \ \ \ \ \ \ (\cos(\omega_0\,x)-1)\cos(\omega_0\,t),0\right]
\ea
\ee
We note that the induced fields are due to the pseudo-photon and stress once more that classically
are obtained by simply considering the standard Maxwell equations.

\subsection{Magnetic Momenta of Fermions}

\begin{figure}[h]
\hspace{1.5cm}\epsfxsize=20mm\epsfbox{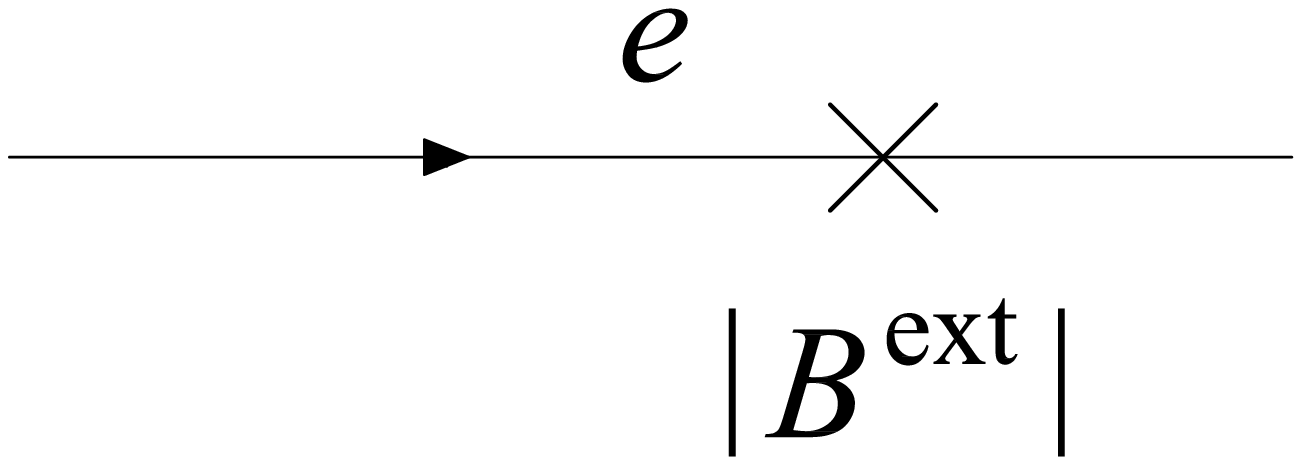}\\[-3mm]

\hspace{1.5cm}\epsfxsize=20mm\epsfbox{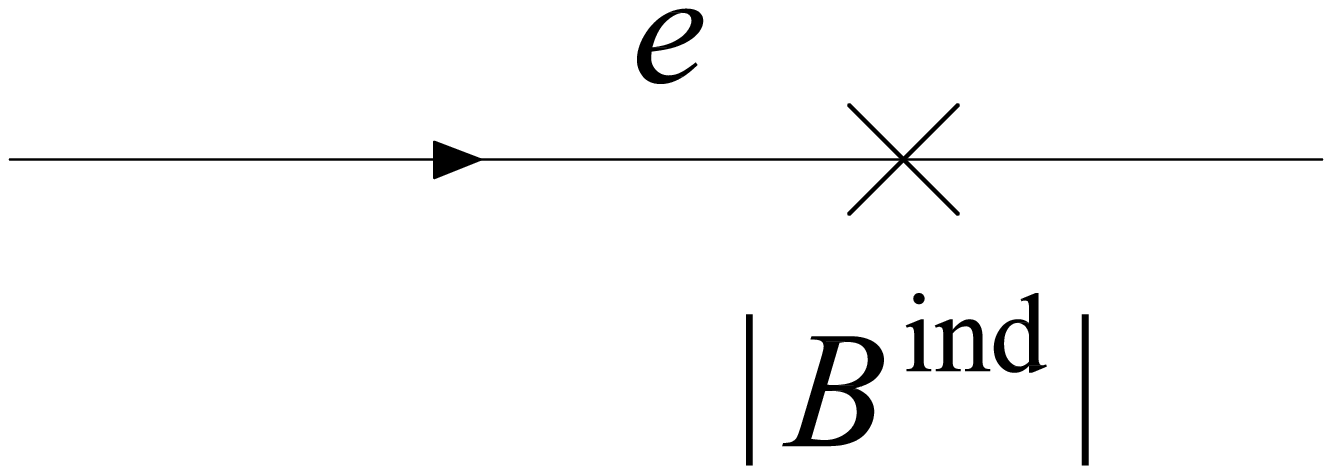}
\caption{\small Diagrams for the magnetic momenta of the electron in the presence of background fields.\label{fig.mm}}
\end{figure}

The magnetic momenta for fermions consist on the effects of the background fields on a free moving fermion
and its magnitude is usually given in terms of the Bohr magneton $\mu_A$. For the case of electrons (see figure~\ref{fig.mm}),
since the induced fields being due to the pseudo-photon we obtain the relative magnitudes
\be
\ba{rcl}
\mu_A&=&\displaystyle\frac{e}{m_e}\\[3mm]
\mu_C&=&\displaystyle\sqrt{\frac{\epsilon_0}{\mu_0}}\frac{g}{m_e}\approx 68.5\times\mu_A
\ea
\ee
Considering an expansion on the radius for the induced magnetic field $|\vb{B}^{\mathrm{ind}}|\approx B_0 \omega_0 r/c$
valid for $\omega_0 r/c<1$ we obtain that the effect of pseudo-photon background becomes more relevant than the effect
of photon background for
\be
\ba{l}
\displaystyle\mu_C\,|\vb{B}^{\mathrm{ind}}|>\mu_A\,B_0\\[3mm]
\displaystyle\Rightarrow\ r>\frac{c}{68.5\,\omega_0}\sim\frac{4.37\times 10^6}{\omega_0}\ meters
\ea
\ee

\subsection{Vacuum Polarization}

\begin{figure}[h]
\hspace{1.5cm}\epsfxsize=20mm\epsfbox{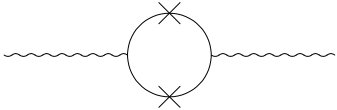}
\caption{\small Diagram for the second order vacuum polarization effects in the presence of background fields.\label{fig.he}}
\end{figure}

The second order vacuum polarization effects are given by the Euler-Heisenberg Lagrangian~\cite{HE} which
accounts for the effect of background electromagnetic fields in the virtual fermion loops (see figure~\ref{fig.he}.
In the presence of both photons and pseudo-photons it reads
\be
\ba{l}
\displaystyle{\mathcal{L}}^{(2)}_e=-\xi_e \left(4({\mathcal{F}}_{\mu\nu}{\mathcal{F}}^{\mu\nu})^2+7(\epsilon^{\mu\nu\delta\rho}{\mathcal{F}}_{\mu\nu}{\mathcal{F}}_{\delta\rho})^2\right)\\[3mm] \displaystyle\xi_e=\frac{2\hbar^3\epsilon_0}{45\,m_e^4c^5}\alpha_e^2
\ea
\ee
with the generalized gauge connection defined as
\be
{\mathcal{F}}^{\mu\nu}=F^{\mu\nu}-\frac{1}{2}\sqrt{\frac{\epsilon_0}{\mu_0}}\,\frac{g}{e}\epsilon^{\mu\nu\delta\rho}G_{\delta\rho}
\ee
For incident radiation of frequency $\omega=ck$ this effect holds a birefringent optical dispersion relation given by
\be
\ba{l}
\displaystyle\omega_\pm=c\,k\,\left(1-\lambda^e_\pm Q_0^2-\lambda^g_\pm Q_{\mathrm{ind}}^2\right)\\[3mm]
\displaystyle\lambda^e_+=7\xi_e\ \ ,\ \ \lambda^e_-=4\xi_e\\[2mm]
\displaystyle\lambda^g_\pm=\frac{\alpha_g^2}{\alpha_e^2}\,\lambda^e_\pm\approx\ 2.2\times 10^7\lambda^e_\pm
\ea
\ee

Considering again an expansion on the radius we have $Q_0^2=B_0^2$ and $Q_{\mathrm{ind}}^2= B_0^2\,\frac{\omega_0^2}{c^2}\,r^2$
such that the effect of pseudo-photon background becomes more relevant than the effect of photon background for
\be
\ba{l}
\displaystyle Q_{\mathrm{ind}}>Q_0\ \Rightarrow\\[3mm]
\displaystyle r>\frac{c}{68.5\,\omega_0}\sim\frac{4.37\times 10^6}{\omega_0}\ meters
\ea
\ee

\section{Conclusion}

We have shown that, although classical extended $U_e(1)\times U_g(1)$ holds the same results of standard
electromagnetism, at quantum field theory level there are measurable effects that can distinguish between
standard QED and a theory containing pseudo-photons. As we have mentioned the simple construction presented
is a conceptual example or toy model. However we stress that this simple example may have relevance in some
particular practical implementations for which magnetic flux tubes or strings are present.

As examples we have type~II superconductivity where vortex solutions exist and, although laboratory electron systems are usually not rotating,
the same solutions may exist in neutron stars and pulsars for which a neutron superconductivity phase could be present~\cite{neutron}.
In stellar plasma (e.g. the Sun) has also been put forward within magneto-hydrodynamics~\cite{Parker} that singular magnetic field
lines may explain the heating of the stellar corona due to magnetic reconnection mechanisms~\cite{Priest}. These results are
also consistent with the field content obtained in planar system for which planar magnetic fields and orthogonal electric fields
exist in pseudo-photon theories (as opposed to the pure Maxwell theory)~\cite{planar}.

As an example we consider neutron stars and magnetars~\cite{magnetars} which hold magnetic fields and rotation frequencies up
to
\be
B_0\sim 10^{12}\ T\ \ ,\ \ \omega_0\sim 10^3\ Hz
\ee
which can have visible effects, for example, in neutrino magnetic momenta measurements~\cite{neutrinos}
and gamma-ray burst polarizations~\cite{polarization}. For the above values we obtain the bounds for
which the effect of pseudo-photon backgrounds are more relevant than the effect of photon backgrounds
of
\be
\ba{rcl}
\displaystyle r_{\mathrm{magn.\ mom.}}&>&4.4\ Km\\[3mm]
\displaystyle r_{\mathrm{vac.\ pol.}}&>&63.9\ m
\ea
\ee
for the magnetic momenta and vacuum polarization effects respectively.

Other frameworks where pseudo-photon effects may be relevant are plasmas and low temperature planar
systems in orthogonal strong electromagnetic fields~\cite{Proca_planar}.\\

\noindent {\bf Acknowledgments} -- 
The author thanks Fernando Barbero, Mar Bastero-Gil, Jo\~ao Pulido,
Jorge Dias de Deus, Pedro Sacramento, D\'ario Passos and Il\'{\i}dio Lopes
for several discussions. This work was supported by SFRH/BPD/17683/2004
and SFRH/BPD/34566/2007.

\end{document}